# Mid Infrared Nonlinear Plasmonics using Germanium Nanoantennas on Silicon Substrates

Marco P. Fischer[1], Aaron Riede[1], Kevin Gallacher[2], Jacopo Frigerio[3], Giovanni Pellegrini[4], Michele Ortolani[5], Douglas J. Paul[2], Giovanni Isella[3], Alfred Leitenstorfer[1], Paolo Biagioni[4], and Daniele Brida[1,*]

[1] Department of Physics and Center for Applied Photonics, University of Konstanz, D-78457 Konstanz, Germany

[2] School of Engineering, University of Glasgow, Rankine Building, Oakfield Avenue, Glasgow, G12 8LT, UK

[3] L-NESS, Dipartimento di Fisica del Politecnico di Milano, Via Anzani 42, 22100 Como, Italy

[4] Dipartimento di Fisica, Politecnico di Milano, Piazza Leonardo da Vinci 32, 20133 Milano, Italy

[5] Department of Physics, Sapienza University of Rome, 00185 Rome, Italy

[*] e-mail: daniele.brida@uni-konstanz.de

**We demonstrate third harmonic generation in plasmonic antennas made of highly doped germanium and designed to be resonant in the mid infrared. Owing to the near-field enhancement, the result is an ultrafast, sub-diffraction, coherent light source tunable between 3 and 5 µm wavelength on a silicon substrate. To observe nonlinearity in this challenging spectral region, a high-power femtosecond laser system equipped with parametric frequency conversion in combination with an all-reflective confocal microscope setup is employed. We show spatially resolved maps of the linear scattering cross section and the nonlinear emission of single isolated antenna structures. A clear third order power dependence as well as the mid-infrared emission spectra prove the nonlinear nature of the light emission. Simulations support the observed resonance length of the double rod antenna and demonstrate that the field enhancement inside the antenna material is responsible for the nonlinear frequency mixing.**

Plasmonic nano-antennas[1,2], i.e. resonant metallic structures with sub-optical-wavelength size, are one of the key components for advanced nano-optics applications. By directing and concentrating the far-field electromagnetic radiation into sub-diffraction-limited near-field volumes, they are the ideal tools to access single quantum systems with light. Another benefit of the light concentration capabilities of resonant antennas consists in the ability to access nonlinear optical phenomena[3] even with minute amplitudes of the electromagnetic fields. This approach has been exploited successfully in the near-IR spectral range to generate second-harmonic[4], third-harmonic[5,6], or even higher-order phenomena[7-9] as well as incoherent, octave-spanning multi-photon photoluminescence[6,10] from gold nano-antennas. Alternatively, mid-IR plasmonic metamaterials have been employed to improve the coupling of far field radiation with the nonlinearity of quantum wells[11].

In recent years, highly-doped semiconductors like InAs[12], InAsSb[13], InP[14] and Ge[15-17] have been introduced as high-quality and tunable mid-IR plasmonic materials for integrated devices. The mid-IR frequency range is of high interest for chemical and biological identification of molecules for environmental, healthcare and security sensing applications, since it includes the so-called molecular fingerprint region[16,18-20] with countless unique vibrational absorption lines. In this context, nonlinear plasmonics can be a sensitive and a versatile tool for achieving a near-field tunable source that can directly interact with molecules at the nanoscale. Up to now, however, intrinsic nonlinearities from plasmonic structures have been demonstrated only in the near-IR and employing standard metals that are not effective at longer wavelengths for this kind of application[21].

For several reasons Ge is the ideal material for linear and nonlinear plasmonic applications in the mid-IR spectral range since, in the last years, extremely heavily-doped materials with a tunable plasma frequency of up to 95 THz (3.1 μm wavelength) and high crystalline quality became available through epitaxy on silicon wafers[17]. The reflectivity and dielectric functions of the material employed in the reported experiments are depicted in Fig. 1a,b. In particular, full CMOS-compatibility of Ge on Si promises easy integration with the potential for cheap mass production. Moreover, the third order nonlinear coefficient[22,23] is comparable to that displayed by gold at visible and near-IR frequencies[24]. Since Ge is a non-polar elementary semiconductor, the lack of optical phonon absorption reduces losses in the mid-IR compared to III-V semiconductors. Finally, the plasma frequency can also be controlled dynamically by optical excitation of electron-hole pairs[25].

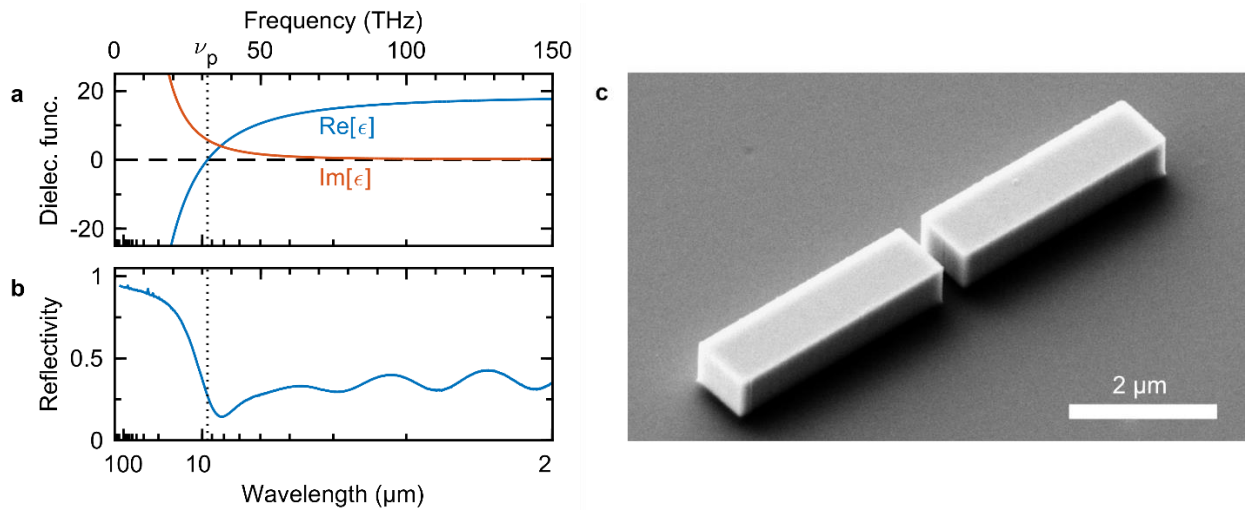

**Figure 1 | Plasmonic Antennas from heavily doped germanium. a**, Complex dielectric function of heavily doped germanium film extracted from **b** the reflectivity spectrum. **c**, Scanning electron micrograph of a single germanium double rod antenna on silicon substrate with an arm length of 3.5 μm and a gap width of 300 nm.

In this work we target the local generation of third harmonic radiation in the mid-IR at sub-wavelength dimensions. Our experiments enable several exciting perspective applications from the nonlinear-enhanced sensing of molecular species[26] to ultrafast near-field microscopy[27,28] with unprecedented temporal and spatial imaging resolution for molecular-selective imaging in life sciences as well as the direct porting of a large variety of near-infrared nonlinear plasmonic applications[3] in the mid infrared. Furthermore the reported technique offers new ways to study the fundamental mechanisms of nonlinearity in condensed matter, especially in nano-structures[24,29].

Nevertheless, the study of nonlinear plasmonics in the mid-IR wavelength range poses major challenges with respect to the near-IR and visible cases. Diffraction, that limits the excitation field in the focus, combines with the increased antenna interaction volume leading to a complex, unfavourable wavelength scaling of the expected third harmonic generation (THG). In addition, the longer pulse durations at low frequencies further decrease the peak fields and thus the nonlinear emission. These difficulties have to be mitigated by scaling the laser system used for the optical excitation. Under these conditions, Ge displays the additional advantage of having a strong durability while in metals strong electro-migration and diffusion prevents high-field excitation for nonlinear mid-IR studies[21].

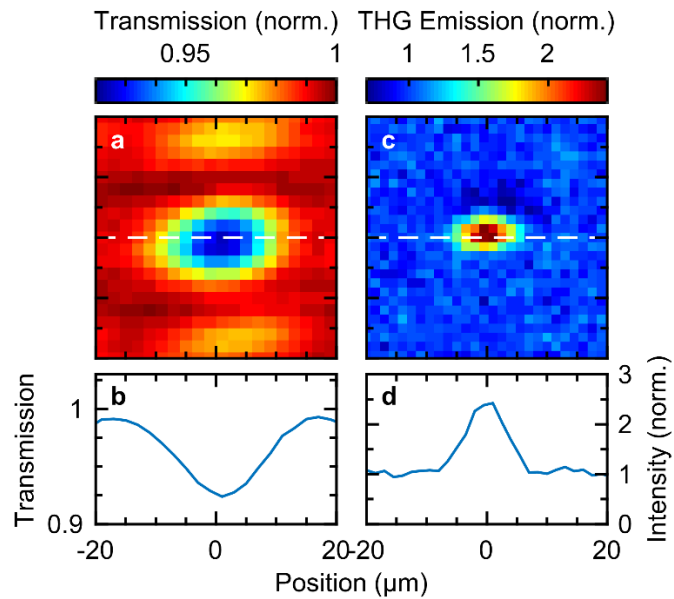

**Figure 2 | Linear and nonlinear confocal microscope images of a single antenna. a**, The spatially resolved transmission map of a heavily doped Ge double rod antenna with an arm length of 4.5 µm at an excitation wavelength of 12 µm recorded by moving the antenna sample through the common objective focus. The darker regions around the antenna are diffraction artefacts of the reflective objectives complex point spread function. **b**, Cut of the transmission map through the center of the antenna (along the white dashed line). **c**, Spatially resolved third harmonic emission intensity from this antenna normalized to the substrate background emission. **d**, Cut through the emission map.

A high-power femtosecond laser system is employed to obtain narrowband, intense, mid-IR transients via difference frequency generation tunable over the wavelength range from 7 to 20 µm. At a pulse energy of up to 100 nJ, the mid-IR driving pulses feature a duration of 300 fs at a bandwidth of 1.5 THz. Single antenna structures (see Fig. 1c) are excited in a dispersion free Cassegrain-geometry reflective microscope setup with these multi-THz transients reaching peak electric fields of up to 5 MV/cm. By scanning the sample through the focal region in a transmission geometry, maps with high spatial resolution can be recorded. Figure 2a demonstrates the transmission map of a double rod antenna with an arm length of 4.5 µm. The excitation wavelength is set to about 12 µm corresponding to 25 THz. The extinction of this single resonant subwavelength structure exceeds 8% (Fig. 2c). Due to the sub-wavelength dimension of the antennas, the geometric width roughly represents the point spread function of the optical excitation. To discriminate between the fundamental and third-harmonic radiation, crystalline filters are employed. This allows the strongly localized third-harmonic emission from the plasmonic structure to be selected (Fig. 2b,d). The emission exceeds the substrate background by a factor of 2.5, which is substantial considering the small interaction volume of the driving beam with the subwavelength structure, in comparison to the bulk silicon wafer interaction path. Both the excitation and the emitted third harmonic generated radiation are polarized along the antenna axis.

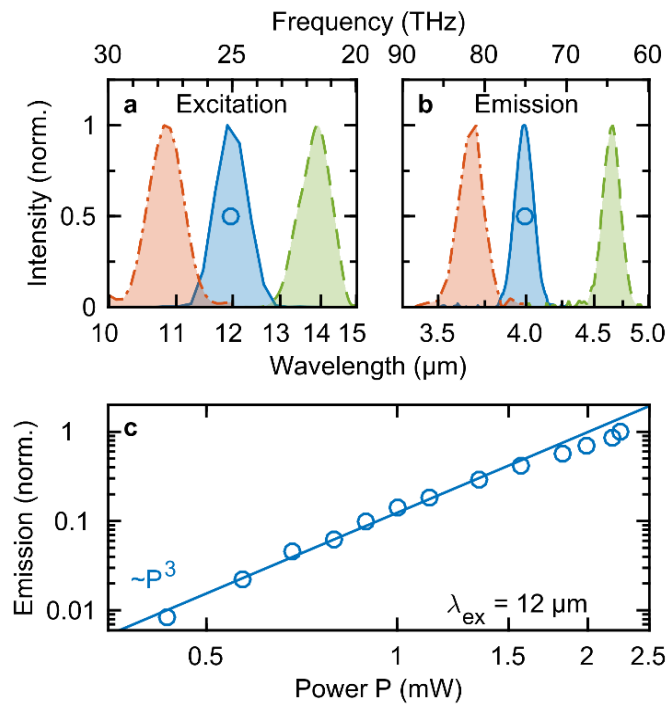

**Figure 3 | Third harmonic emission spectra and nonlinear power dependence. a**, The excitation spectra and **b** the corresponding third harmonic emission spectra from single resonant double rod antennas at excitation wavelengths of 10.7 µm, 12 µm and 14 µm (red, blue and green, respectively). **c**, The nonlinear power dependence of the THG emission in double-logarithmic scale for 12 µm excitation wavelength (circles). The line of cubic proportionality proves the third harmonic character of the radiation and shows the start of saturating behavior for high excitation powers beyond 1.5 mW.

Figures 3a,b show the exact emission spectrum under three different excitation conditions. The power dependence of the emitted THG radiation from the antenna centre (Fig. 3c) with a cubic power exponent further proves that we indeed observe third-harmonic emission in this experiment. At high pump intensity levels, the curve demonstrates that the efficiency of the nonlinear emission starts to deviate from a third order power law. This could be caused by transient heating of the antenna that modifies the dielectric behaviour of Ge or by charge carrier excitations that slightly increases the reflection losses.

Taking into account the losses of the optical elements employed for the collection of the third harmonic, we estimate a yield of the nonlinear frequency conversion (defined as the number of third-harmonic photons generated per excitation photon) that is higher than $10^{-6}$ for a single antenna emitter illuminated by 25 nJ driving pulses. This calculation does not include the limited collection aperture of the condenser objective with respect to the emission pattern of the THG. If a standard dipole emitter coupled to the antenna gap is considered to mimic the third-harmonic radiation, the total energy conversion ratio can be estimated to be approximately $10^{-5}$ with a THG energy in the order of 1 pJ per pulse.

Finally, we investigate the dependence of both the linear scattering cross section and the third harmonic emission on the antenna arm length. Extinction and emission data are extracted from confocal microscopy raster scans similar to Figure 2 for a set of different antennas. The arm length is varied between 1.0 µm and 6.0 µm at a constant gap width of 300 nm. Figure 4a demonstrates the measurement results for an excitation wavelength of 12 µm. By normalizing over the antenna volume, a clear resonance for an arm length of 3.5 µm becomes visible. As expected from the nonlinear scaling of the excitation, this effect becomes more prominent in the THG emission. For antennas below 1.5 µm arm length, no third harmonic could be detected, whereas antenna arms longer than 3.5 µm become less efficient because of off-resonance conditions. The measurement is repeated in Figure 4c for an excitation wavelength of 14 µm displaying a shift of the resonance to longer antennas of approximately 5.0 µm arm length.

Figures 4b,d display the results of three-dimensional finite-difference time-domain simulations that reproduce the observed behaviour with high fidelity, taking into account the numerical aperture of the two objectives and evaluating the THG signal by considering the third power of the field intensity integrated over the whole antenna volume. As previous studies revealed, the relatively large thickness of the antenna arms leads to two distinct resonances at the Ge-Si interface and at the Ge-air interface[16]. The latter of these resonances is excited in this experiment.

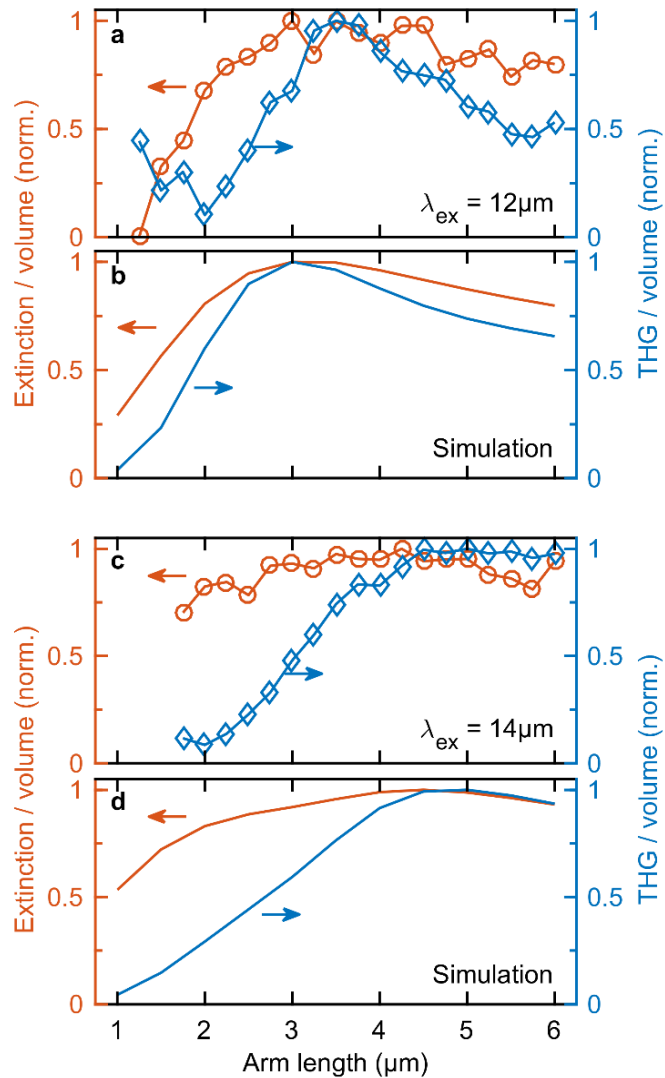

**Figure 4 | Measurement and simulation of extinction and THG emission for antennas of various arm lengths. a,c**, Normalized extinction (red circles) and third harmonic emission (blue diamonds) per unit antenna volume measured for double rod antennas of various lengths at a gap width of 300 nm excited at 12 µm and 14 µm central wavelength, respectively. **b,d**, Corresponding FTDT simulation results for both excitation conditions.

In conclusion, we were able to demonstrate for the first time the coherent nonlinear emission from plasmonic subwavelength-structures in the mid-IR frequency range. This achievement is enabled by the crucial advancements in the growth of epitaxial group-IV semiconductors and of heavily-doped germanium on silicon substrates in particular. Our results pave the way for new methods in mid-IR near-field microscopy employing germanium nanostructures[30] and for the sensing of molecules based on their vibrational absorption fingerprints. This capability combined with the integration in a CMOS platform will allow targeting sensitive applications in biomedicine, international security for the detection of hazardous compounds and even environmental protection for example by monitoring the emissions of combustion engines. In addition, semiconductor plasmonics is promising for fundamental studies of nonlinear light-matter interaction. In perspective, new insight into controversial questions about the origin of nonlinear susceptibility in nanostructures[24,29] and its spectral dispersion[6] can be gained by tuning the plasma edge of doped germanium thus providing a degree of freedom not accessible with standard metals. Finally, combined with the all-optical ultrafast generation of free carriers in intrinsic Ge nanostructures[25], this work opens the possibility for active control over coherent mid-infrared light sources with unprecedented spatiotemporal confinement.

**Methods:**

**Heavily doped Ge film growth and material characterization.** As a first fabrication step, a highly n-doped Ge epilayer was grown on an intrinsic Si(001) substrate by low-energy plasma-enhanced chemical vapour deposition (LEPECVD)[31]. The deposition was performed at 500 °C with a growth rate of about 1 nm/s. The n-type doping was obtained in-situ by adding 0.035 standard cubic centimetre per minute (sccm) of $PH_3$ to 20 sccm of $GeH_4$, achieving an active doping of $2.5×10^{19}$ $cm^{-3}$, which corresponds to a plasma frequency of about 31 THz (9.7 µm wavelength). The thin film reflectivity spectrum is recorded using Fourier-transform infrared spectroscopy and allows extracting the dielectric function of the material (Fig. 1a,b) by a combination of Drude-like modeling and the use of the Kramers-Kronig relations[17]. It should be noted here that recent works also demonstrated that the same growth technique, when combined with post-growth annealing procedures, allows plasma frequencies up to about 95 THz (3.1 µm wavelength) to be reached[32,33].

**Antenna fabrication.** From these heavily doped Ge films, isolated antenna structures (Fig. 1c) are fabricated via electron-beam lithography with hydrogen silsesquioxane resist and anisotropic reactive ion etching with fluorine chemistry[34,35]. The half-wavelength length of the two arms in the double-rod antenna geometry is selected in order to maximize the resonantly-enhanced currents. Considering the refractive index of the material and of the environment, numerical simulations predict that the first-order localized plasmon-polariton resonance occurs for an arm length of 3 to 4 µm when excited at a wavelength around 12 µm with linearly polarized light. By the use of state-of-the-art nanofabrication technology, the relatively large structures can be fabricated close to perfection with steep sidewalls, sharp edges and reproducibility to within a nm of the design[34].

**Mid-IR laser microscopy setup and antenna characterization** To excite the resonant antenna structures, we developed a high-power femtosecond laser system[36] based on a regenerative Yb:KGW amplifier with a repetition rate of 50 kHz. The fundamental pulses at a wavelength of 1028 nm with a duration of 250 fs drive a non-collinear optical parametric amplifier (NOPA) [37,38] that generates broadband pulses tunable between 1050 and 1400 nm. Intense mid-IR pulses are subsequently produced via nonlinear phase-matched difference frequency mixing of the NOPA pulses (pulse energy up to 2.2 µJ) with 36-µJ-pulses from the Yb:KGW amplifier in a GaSe crystal[39]. Depending on the crystal thickness, phase-matching angle and NOPA wavelength, either broadband mid-IR transients spanning from 8 to 20 µm wavelength or narrowband but intense pulses of up to 100 nJ tunable over the same spectral range can be generated. For these experiment, driving mid-IR pulses are set to a duration of 300 fs at a bandwidth of 1.5 THz.

A custom-build confocal microscope consisting of two dispersion-free gold-coated reflective Cassegrain objectives with a numerical aperture of 0.5 and working in transmission geometry is used to study the linear and nonlinear mid-IR response of individual antenna structures. The first objective focusses the driving pulses onto the antenna samples yielding excitation fields of up to 5 MV/cm and reaching a diffraction limited spot size of approximately one wavelength. The second objective collects the transmitted radiation as well as the nonlinear emission. A pinhole with a diameter of 150 µm in the image plane reduces the detection field of view to enhance the lateral resolution, rejects stray light coming from outside the focal volume and lowers the depth of field to better discriminate between the substrate bulk emission background and the antenna signal. The detection of the transmitted and emitted nonlinear radiation is performed via electro-optic sampling (EOS) in 80-µm-thick GaSe[40] or with a liquid nitrogen cooled mercury cadmium telluride (MCT) detector with low-noise lock-in readout. To discriminate between the fundamental and third-harmonic radiation, crystalline filters of InSb (a long-pass filter with a transmission edge at 7 µm wavelength), sapphire (a 5 µm short-pass filter), and $CaF_2$ (a 9 µm short-pass filter) are employed. A monochromator with blazed mid-IR gratings was employed to record the exact emission spectrum from the samples.

**Simulation of the antenna response.** Three-dimensional finite-difference time-domain simulations (*FDTD solutions, Lumerical Inc.*[41]) were performed to reproduce the observed extinction and emission spectra. The dielectric constant of the heavily-doped Ge material and of the underlying Si substrate was experimentally characterized as described above and used for the simulations. The antennas are illuminated with a Gaussian source at the fundamental wavelength, featuring the same numerical aperture as the illumination objective in the experimental setup. The influence of the numerical aperture of the collection objective in the extinction maps was taken into account as well by considering the radiation pattern of the antennas at the fundamental wavelength. In order to model the nonlinear signal and reproduce its dependence on the arm length, we calculate the third power of the electric field intensity integrated over the whole antenna volume, mimicking a standard third-order bulk nonlinear process inside the antennas. This simple approach already reproduces the experimental data with high fidelity.

**Acknowledgements**

The authors acknowledge support from European Commission via the Marie Curie Carrier Integration Grant and the Seventh Framework Programme under Grant No. 613055, from the Deutsche Forschungsgemeinschaft through the Emmy Noether programme and EPSRC under Grant No. EP/N003225/1. The authors would like to thank the staff of the James Watt Nanofabrication Centre for help with fabricating the devices.